# Nonlinear Dynamics in Psychophysiology - Importance of Time Scales

**Wlodzimierz KLONOWSKI**

*Lab. of Biosignal Analysis Fundamentals*
*Institute of Biocybernetics and Biomedical Eng., Polish Academy of Science,*
*4 Trojdena St., 02-109 Warsaw*
*POLAND*
<u>wklon@hrabia.ibib.waw.pl</u> , <u>http://www.ibib.waw.pl/~lbaf</u>

Presented theory of feelings and emotions is based on Nonlinear Dynamics and Theory of Complex Systems. The most important assumption is that the brain may be considered to be composed of subsystems characterized by different characteristic time scales. The theory enables to define feelings in a physicalistic way. It also explains a possible role of feelings and emotions in cognition. We propose to call presented theory *Chaosensology*.

*"Nur in der Sprache der Mathematik kann die Welt angemessen beschrieben werden."*
*Ja - aber wie wenig Menschen sind imstande, die herbe Schönheit einer solchen*
*Beschreibung zu verstehen!*

Esther Vilar "Die Mathemaitk der Nina Gluckstein"

*"Only in the language of Mathematics the world can be adequately described."*
*Yes – but how few people are able to understand the harsh beauty of such a description!*

Esther Vilar "The Mathematics of Nina Gluckstein" (transl. from German W. Klonowski)

# 1. Introduction

I am a Physicist, and Physicists use to think in an exact logical manner. And maybe that is why I have always been fascinated about *feelings* as 'something that seems to have no logical explanation'. This statement became for me a kind of a definition of the notion of 'feelings' (Lat. *sensus*). In dictionaries one can read e.g. 'Feelings – emotional side of a person's nature (contrasted with intellect): The speaker appealed to the feelings of the audience rather than to their reason' [1]; 'Emotion – strong feeling of any kind: love, joy, hate, fear and grief are feelings' [1].

Physicalistic theories may be applied to different philosophical concepts (cf. [2]). A long time ago I became discouraged to read books and papers dealing with Psychology because most of them have been rather inconsistent logically and different notions have been terribly confused. When I started work on biocybernetics of the brain, trying to apply methods of Nonlinear Dynamics [3-5], I also began to think how to define feelings more precisely and what is a role of feelings in cognition.

Our world is not governed by simple linear laws we were taught in schools. If for one man it takes one hour to dig a ditch it does not mean that for 60 men it will take just one minute to dig the same ditch – this is obvious for everybody, even for those who do not know contemporary Physics and Mathematics. What in Physics is called *nonlinearity* it is just stating the fact that elements forming a system do not act independently one from another. In the example above the system consists of one ditch and 60 men – the men may not dig independently from one another simply because of the limited size of the ditch. In Medicine it is well known that the effect of a drug is not just simply proportional to its dose; also mutual interaction of drugs is extremely important. 'In Medicine two plus two often does not make four' [6] and a small dose of a drug may act the other way than a greater dose; for example, alcohol, caffeine, nicotine in small doses act as stimulants but in greater doses are just poisons [7].

Nonlinear Dynamics is *interdisciplinary*. We all become more and more specialized in some very narrow disciplines and we often do not know that the methods we want to apply in our research have been used for a long time in other disciplines. When we learn about it we are often amazed like Molier's Mr. Jourdain ("Le Bourgeois Gentilhomme" II.iv) who says: 'Good heaven! For more than forty years I have been speaking prose without knowing it.' The most likely reasons for failing to observe some phenomena and processes is that we often don't see what we don't look for.

## 2. Concept of time scale

It is not an exaggeration to say that in Physics everything is a question of *time scale*. Disintegration of radioactive isotopes may serve as a good example. In classical Physics time is considered to be a continuous variable and processes are modeled using differential equations. Radioactive disintegration is modeled by a simple ordinary differential equation

$$dz/dt = -a \cdot z \tag{1}$$

This equation simply means that process rate, i.e. the number of atoms that disintegrates in a unit of time (e.g. 1 sec.), $dz/dt$, is proportional to the number of atoms, $z$, present in the system at a given moment $t$; minus sign means that the total number of atoms of the given isotope diminishes with time; $a$ is a constant.

Solution of equation (1) has the form

$$z(t) = z_0 \cdot exp(-t/\tau) \tag{2}$$

where $z(t)$ is the number of atoms at moment $t$, $z_0$ - number of atoms at the initial moment $t=0$, and $\tau$ is *characteristic time of the process*, a kind of a natural time unit that is an intrinsic characteristic of the considered process – the *greater* is $\tau$ the smaller number of these units is in one second (minute, hour, …), and so the *slower* is the process. Characteristic time $\tau$ is reciprocal of the proportionality constant $a$ from equation (1)

$$\tau = 1/a \tag{3}$$

and it is related to so called disintegration *half-time*, $\tau_{1/2}$, i.e. time needed for disintegration of one half of the total number of atoms present in the system at the given moment

$$\tau = \tau_{1/2} / ln2 \tag{4}$$

Let us consider a system containing atoms (isotopes) with different disintegration half-times. And let's say we are interested how the system will look after 1 hour. We may drastically simplify description of the system by classifying atoms that are present in the system into three classes:

1. those changing quickly, i.e. their characteristic times are much shorter than the period of observation (in the above example those with $\tau_{1/2}$ from less than a second to 5 minutes);

2. those changing slowly, i.e. their characteristic times are comparable with the period of observation (in the above example those with $\tau_{1/2}$ from 5 minutes to 12 hours);

3. those changing very slowly (in the above example those with $\tau_{1/2}$ greater than 12 hours).

It is obvious that during 1 hour total number of atoms belonging to class 3. basically will not change, while almost all atoms belonging to class 1. will disintegrate; so, we need to solve only differential equations concerning the atoms of class 2.

In general case, *quick variables* (those for which characteristic times are much shorter than the period of observation we are interested in) reach their *stationary values*, while *very slow variables* (those for which characteristic times are much longer than the period of observation) stay constant and may be treated just as parameters. We need to solve only equations for *slow variables* (those for which characteristic times are comparable with the period of observation). It is necessary to emphasize that it is relative (it depends on the period of observation we are interested in) which variables are considered to be quick and which are considered to be slow. There exist mathematical theories, e.g. Tikhonov's Theorem, that enable elimination of quick variables in system modelling (cf. [8]).

## 3. Phase space and the brain

In contemporary Physics instead of continuous time $t$ one often uses *discrete time*, $t_i$, and instead of continuous functions $z(t)$ one often uses *discrete maps* that transform values $z_t$ of state variables at a given moment, $t_i$, onto values of these variables $z_{t+1}$ in a next moment, $t_{i+1}$, i.e. one discrete unit of time later. Any *dynamical system*, no matter how complicated, may be depicted using a multidimensional *phase space (state space)* with coordinates chosen in such a way that a point in that space entirely characterizes *the state of the system* at a given moment and *dynamical rules* that depict evolution of all state variables, i.e. determine the state of the system in a subsequent moment $t_{i+1}$ if the state at the moment $t_i$ is known. Any change of the system's state is called a *process*. So, the phase space is the set of all possible states of the system, while a *trajectory* in the phase space depicts evolution of the system with time, i.e. processes that take place in the system. Already at the turning of XIX and XX centuries H.Poincaré demonstrated by using notion of phase space that systems can behave very differently than it has been predicted by classical Newtonian dynamics, for example that so called *deterministic chaos* i.e. extreme sensitivity to initial conditions is possible

*Nonlinear Dynamics* – a branch of Complex System Theory - where the effect of a stimulus is not proportional to the strength of that stimulus, introduces the notion of *attractors* that is such subsets of system's phase space that trajectories of the system tend ('are attracted') to those subsets. For living systems attractors correspond to states of *homeostasis*. Chaotic systems have so called *strange attractors*, characterized by a non-integer dimension.

Dynamics of a living system may be represented as a trajectory in a multi-dimensional phase space with the axes measuring e.g. concentrations of different enzymes, blood flow, rates of recurrence of meeting other organisms of the same and of different species, intensity of different forms of behavior, etc.; such a system may be treated as being composed of some minimal *subsystems*, such that further subdivision of those subsystems does not make sense – they are represented by some subsets of the system's phase space that interact one with another only in a very weak manner [9].

Brain is a complex dynamical system that can show some characteristics of deterministic chaos. *Brain phase space*, $\mathcal{B}$, may be defined as follows:

$$\mathcal{B} = (\vartheta, \mathcal{M}, \mathcal{R}, \mathcal{E}, \mathcal{C}, \mathcal{O}) \tag{5}$$

where $\vartheta$ denotes the set of *input variables* i.e. those describing influence of the environment and of the body on the brain (e.g. impulses from sense organs), $\mathcal{M}$ denotes the set of *memory* variables, $\mathcal{R}$ denotes the set of *operational memory* variables, $\mathcal{E}$ is the set of variables defining the *emotional state*, $\mathcal{C}$ is the set of variables defining the *state of consciousness*, $\mathcal{O}$ is the set of *output variables* i.e. those describing influence of the brain on effectors (e.g. the nerve impulses that are transmitted to muscles).

Contemporary knowledge of brain physiology and biochemical physics is not sufficient to point out which specific variables belong to each of individual subset of the phase space $\mathcal{B}$, in particular to subsets $\mathcal{E}$ and $\mathcal{C}$. While considering different processes in the brain I explicate below the following **hypothesis – state of consciousness is defined by slow variables, emotional state is defined by quick variables** (cf. above remarks concerning notion of characteristic time scales).

### 4. Feelings and Emotions

Dynamical rules that govern changes of brain state have not yet been scientifically understood. Only abstract description of such rules is possible. It is not easy even to imagine that these rules will ever be recognized in details since brain phase space undoubtedly has very many dimensions and only in very specific states the number of essential dimensions may really be reduced. Reduced dimensionality probably correspond to pathological brain states. For example, analysis of electroencephalographic signals, EEG, using methods of Nonlinear Dynamics shows, that during epileptic seizure brain acts in a more ordered fashion than a healthy brain. It is also well known that even simple nonlinear dynamical rules may present very complicated 'behavior'. For example, simple logistic map that maps a unit interval $U$ onto itself:

$$f: U \to U \quad \text{where} \quad U = [0, 1] \qquad (6a)$$

$$z_{t+1} = r \cdot z_t \cdot (1 - z_t) \qquad (z_t \in U) \qquad (6b)$$

shows very different „behavior" depending on the value of parameter $r \in [0, 4]$ - it changes from a simple point attractor (for $r \in [0, 3]$), through periodical attractors characterized by different periods (for $r \in [3, 3,85]$), up to complete chaos with periodical windows (for $r \in [3,85, 4]$) [3].

In the case of dynamical rules governing the brain, not only any state variable may be subjected to different dynamical rules but also parameters of such a map may change with time (i.e. they may themselves depend on the actual state of the system) or the maps remain unchanged but the whole phase space of the system changes continually. Mathematical theory that would make analysis of such systems possible does not exist yet.

That is why we need to use the notion of time scale to subdivide the variables describing state of the brain into classes. Let us assume that we are interested in brain activity in such a time scale that the parameters of the dynamical maps that transform brain state variables remain (at least in the first approximation) unchanged. For example, we assume that the number of neurons (nerve cells) in the brain remains constant, since the time scale characterizing processes that change the number of neurons is much greater than the time scales that characterize changes in connections between neurons (synapses), changes in concentration of special chemicals called neurotransmitters, and of course changes of the "input" (i.e. of the variables $\vartheta$).

In general, time evolution of the state of brain is given by a map

$$\mathcal{F}: \mathcal{B}_t \to \mathcal{B}_{t+1} \qquad (7)$$

where index $t$ denotes the given moment and $1$ denotes the assumed unit of time, characteristic for processes to which the model is applicable.

According to our hypothesis, characteristic time scale of changes of states of consciousness, defined by slow variables, is much longer than that of emotional states, defined by quick variables. So, several emotional states may have changed before the state of consciousness does change, i.e. there exist processes that change emotional state without changing state of consciousness (for simplicity we assume that input, output, and memory variables also remain unchanged)

$$\mathcal{F}_{\mathcal{E}} : \mathcal{E}_t \rightarrow \mathcal{E}_{t+\tau} \tag{8}$$

where $\tau$ denotes the unit of time scale of quick variables; such processes we will call *emotional processes* or *feelings*.

Similarly, processes that change the state of consciousness

$$\mathcal{F}_{\mathcal{C}} : \mathcal{C}_t \rightarrow \mathcal{C}_{t+T} \tag{9}$$

where *T* denotes the unit of time scale of slow variables (*T* is much greater than $\tau$), we will call *thinking processes* or *thoughts*.

Processes that cause a change of the state of consciousness at the moment *t+T* depending on the emotional state at the moment *t*

$$\mathcal{F}_{\mathcal{A}} : \mathcal{E}_t \rightarrow \mathcal{C}_{t+T} \tag{10}$$

we will call *awareness processes*. Characteristic time scale of awareness processes is much greater than that of emotional processes; for simplicity we assume that it is equal to characteristic time scale of reasoning processes, *T*.

Processes that change the state of consciousness, i.e. thinking processes and awareness processes, we shall also call *cognitive processes*. So, the difference between emotional processes and cognitive processes lies in the difference of characteristic time scales. In the short time scale emotional processes change *structure of the brain phase space*, $\mathcal{B}$ (at least locally), reaching a kind of a stationary emotional state; cognitive processes act on such a phase space that has already changed its geometry (structure) due to emotional processes. This way some parts of the phase space become more easily reachable. This may correspond to the well-known psychological phenomenon of 'thoughts being concentrated on the object of feelings'.

Any change of emotional state (due e.g. to a stimuli reaching the brain from senses or from internal receptors, from memory or as a feedback from output variables) is, according to our definitions, a kind of feeling. When a stimulus changes emotional state then after a sufficiently long time the state of consciousness may eventually also be changed – the subject becomes *aware of the feeling*. It concerns also so called *higher feelings*. For example, falling in love in 'an object that met the eyes' does happen as quickly as an involuntary *reflex* of hand withdrawal when one touches a very hot surface – only after a while we become aware that the surface we touched was really hot; similarly, what does reach the consciousness is not the *emotion of falling in love* but *awareness* of this emotion, but then is already too late…

Of course, it does not mean that emotional processes (feelings) are completely independent of thinking processes (thoughts). In a sufficiently long time scale thoughts do affect feelings. Even if thoughts do not affect directly the emotional state they certainly affect memory $\mathcal{M}$ and output variables $\mathcal{O}$, that in turn through feedbacks affect input variables $\mathcal{I}$, operational memory $\mathcal{R}$, and emotional variables $\mathcal{E}$ - if we are aware of something then our emotions concerning that are very different than if we were not aware of it. In such a way cognitive and emotional processes, thoughts and feelings, influence each other.

**5. Chaos of feelings**

The brain may neither be fully deterministic because in such a case creating of new ideas would be impossible – everything would be determined in advance, like in a case of throwing a stone the distance it reaches before touching ground is exactly determined by the velocity and the angle at the moment of throwing (if air resistance may be neglected); nor the brain may be fully stochastic, because then we would not be able to perform any activity with intended purpose like e.g. touching with our finger the given key on the keyboard.

This dilemma may be solved by the notion of *deterministic chaos* as introduced in Nonlinear Dynamics ([3]-[5]). The laws that govern behavior of a *chaotic system* are deterministic, and its state in any moment would have been completely determined by the state at the initial moment if only the initial state and system parameters had been given with infinite precision. Chaotic system, unlike 'classical' deterministic system, is extremely sensitive to initial conditions and system parameters – even an extremely tiny change in the initial moment causes a huge difference after sufficiently long time – the system reaches completely different state.

In the case of *creative processes* (*higher nervous activity* and *executive functions*) brain behaves like a chaotic system. It means that during such a creative process trajectories in the phase space of the brain tend to certain limited regions (attractors). The brain does not perform searching in the whole phase space but only in a small subset of it. It may be compared to a searching on Internet using a search engine like Google. If we write in the search box term 'brain' the search engine will find so many www pages and other sources that our whole life, 24 hours a day 7 days a week, will not be enough to have even short look through all of them. Moreover, the volume of sources that will appear on Internet in any period of our studies will probably be greater than the volume of sources we look through during the same period. So, even if we had lived *infinitely long* we would have never succeeded to look through all the sources concerning 'brain'. It is a good example what *infinity* really is. If we liked to look through the sources on Internet in a reasonable time we would need to make our search much more constricted.

While making a search on Intenet we often find an item we have not been looking for, we have not even known it exists anywhere. In some sense we *discover* something, we make an *invention.* We also know that making a small mistake in the word we write in the search box, like for example the lack of the dot in ,i' in the word 'brain' would cause that either no sources would be found or the very different ones (such a tiny mistake is not possible in English, but for example Polish word for brain is 'mózg' with a tiny acute over 'o' , and the word 'mozg' does not exist in Polish, but it does mean brain in Croatian; in English we may imagine, for example, writing in the box by mistake 'train' or 'rain' instead of 'brain'). Similarly, feelings exert influence on cognitive processes by changing geometry of the phase space of brain and so 'initial condition' of thoughts; such phenomena has often be called *cognitive intuition.*

Geometry of phase space of the brain is non-planar and non-Euclidean. Some emotional states that may seem to us very different or even in some sense opposite may in the phase space be represented by regions that are close or even adjacent one to another, like on the Earth Far East is adjacent to Far West along the date-change line. If we start moving thousands kilometres towards west from Warsaw we find ourselves in Far West; but if we start from Lodz (about 100 km west of Warsaw) instead, then after passing the same distance towards west we can suddenly find ourselves in Far East instead in Far West. There exists a well-known psychological phenomenon that feelings are often transformed into exactly opposite ones, for example love is transformed into hate; in our theory that we call *chaos of feelings* a tiny change of 'initial conditions' is enough for such transformation to happen.

## 6. Consciousness and subconsciousness

The ideas presented above are verified by the newest research in neuropsychology. Neuropsychologists observed that what had ben considered to be a 'product of consciousness' often occurs so quickly that we have not enough time to think, that we do act first and only later we become aware of what we had just done [10]. For example, sometimes we get aware that we had said something 'without a thought'. We may observe that some persons often speak out first and only then they eventually think what they did say. Neurophysiologists more and more often bring forward arguments that it concerns all of us, that we all speak first and only later we somehow become aware of what we had said, and that it is like this practically with any human activity [11] - speaking, writing, and all informational processes in the brain would take place 'on Level 2' that is in *subconsciousness*. and only some of those, chosen by some kind of a 'decision-making device' called *Central Executive Structure, CES*, would be eventually able to reach 'Level 1', i.e. *consciousness*. So, our conviction that we are in control of our consciousness would be just a delusion [12].

While the concept of 'two levels' may easily be accepted, it is rather impossible for a Physicist to accept the concept of a Central Executive Structure. For sure, no special neuro-anatomical structure that may act like this has ever been found. And question arises what would be the bases of such a CES action as a 'border control' that decides what can pass from Level 2 to Level 1 and what cannot, i.e. which feeling may 'reach consciousness'.

Our physicalistic theory of feelings gives simple explanations of these problems. According to our hypothesis 'Level 1' (*consciousness*) consists of states $\mathcal{C}$, while 'Level 2' (*subconsciousness*) consists of emotional states, $\mathcal{E}$. Due to the big differences in characteristic time scales for emotional processes (8) and for cognitive processes ((9) and (10)), during time $T$ that is necessary for the subject to become aware of some feeling, many other emotional processes may take place, changing phase space of the brain $\mathcal{B}$ (even if very slightly), so causing that because of the 'chaos of feelings' the awareness process of that feeling (cp. (10)) does not occur at all. So, any CES, any decision-making device to control the border between subconsciousness and consciousness, is not needed.

## 7. Conclusions

Despite of the fact that we took love feelings as an example, if one accepts the definition of feelings (8) then any 'recognized object' causes some feelings i.e. changes emotional state of the brain. In quick time scale a feeling changes brain phase space and in such a way it limits 'wandering' on this phase space, so 'giving directions' to cognitive processes

Further evolving of the presented theory should take into account the role played by Memory (variables from the set $\mathcal{M}$ in (5) that may still be subdivided into variables that characterize *long-term memory* and variables characterizing *short-term memory*) - the competition between perception in the given moment and recollection of the past states [12], as well as the role played by a limited capacity of the Operational Memory (variables from the set $\mathcal{R}$ in (5)) that may explain several phenomena observed by psychologists [13].

One should always remember about possible feedbacks. Cognitive processes also change phase space of the brain, what in turn may influence (after sufficiently long time) emotional processes i.e. consciousness may influence subconsciousness. Some authors (cf. [11]) forget about possible feedbacks - their CSE acts like a diode – it lets 'pass through' only in one direction. Neglecting feedbacks, as shown by history of science, may lead to caricatural result. For example, in so called "Political Economy of Socialism" according to Oscar Lange (a famous Polish economist, a founder of Econometrics, author of interesting works concerning Statistics and Cybernetics) 'base' actively influenced 'superstructure' but not oppositely – development of the 'superstructure' has not any influence on 'base'; in practice, it did mean that one should invest in the 'base', mainly in heavy industry, while investments in 'superstructure', i.e. in cultural and scientific development, were considered to be of no economic importance. The results of such economical doctrine we still bear over us in Poland.

Recapitulating: Nonlinear Dynamics may be applied in Psychophysiology of Feelings and Thoughts and it may help in understanding differences between cognitive and emotional processes. It is true what Esther Vilar says: *"only in the language of Mathematics the world can be adequately described"*. We propose to call presented theory **Chaosensology**.

**References**

[1] *The Advanced Learners Dictionary of Current English* by A.S.Hornby, E.V.Gatenby, H.Wakenfield, Seconf Edition, Oxford University Press, London, 1963;

[2] Klonowski W.: Omnis Felicitas Ex Felicitate: Physicalistic Concept of Happiness, in: *Uroboros or biology between mythology and philosophy*, W.Lugowski and K.Matsuno, Eds., Arboretum, Wroclaw, pp. 315-320, 1998;

[3] Stewart I.: *Does God Play Dice? The New Mathematics of Chaos,* Penguin Books, 1990;

[4] Tempczyk M.: *World of harmony and world of chaos* (in Polish), PIW, Warsaw, 1995;

[5] Kapitaniak T., Wojewoda J.: *Bifurcations and Chaos* (in Polish), Lodz Technical University and Polish Scientific Publishers (PWN), Warsaw – Lodz, 2000;

[6] Wiland-Żera A.: personal communication.

[7] Klonowski W.: Non-Linearity and Statistics - Implications of Hormesis on Dose-Response Analysis, *Biocyb.Biomed.Eng.* **19**, pp. 41-55, 1999;

[8] Klonowski W.: Simplifying Principles for Chemical and Enzyme Reaction Kinetics, *Biophys.Chem.* **18**, pp. 73-87, 1983;

[9] Emlen J.M, Freeman D.C., Mills A., and Graham J.H.: How organisms do the right thing: The attractor hypothesis, *Chaos* **8**(3), pp. 717-726, 1998;

[10] Damasio A.R.: *Descartes' Error: Emotion, Reason and the Human Brain*, Papermac, 1994;

[11] Halligan P.W., and Oakley D.A.: Greatest Myth of All, *New Scientist*, No. 2265, pp. 34-39, 18 Nov. 2000;

[12] Dennett D.: *Consciousness Explained*, Penguin Books, 1991;

[13] Brooks M.: Fooled Again, *New Scientist*, No. 2268, pp. 24-28, 9 Dec. 2000.